\documentclass{eas}
\usepackage{graphicx}
\begin{document}

%%-----------------------------
%%      the top matter
%%-----------------------------
\title{Formation and evolution of dwarf galaxies in the CDM Universe} 
\runningtitle{Formation and evolution of dwarfs in the CDM}
\author{Lucio Mayer}\address{Institute fir Theoretical Physics, University of Z\"urich, 8057, Winterthurestrasse 190, Z\"urich}
%\author{...}\address{...}
%\author{...}\address{...}
%
%
\begin{abstract}

We first review the results of the tidal stirring model for the transformation of gas-rich dwarf 
irregulars into dwarf spheroidals, which turns rotationally supported stellar systems into 
pressure supported ones. We emphasize the importance of the combined effect of ram 
pressure stripping and heating from the cosmic ultraviolet background in removing the gas
and converting the object into a gas poor system as dSphs. We discuss how the timing of 
infall of dwarfs  into the primary halo determines the final mass-to-light ratio and 
star formation history. Secondly we review the results of recent cosmological simulations of
the formation of gas-rich dwarfs. These simulations are finally capable to produce a realistic
object with no bulge, an exponential profile and a slowly rising rotation curve. The result
owes to the inclusion of an inhomogeneous ISM and a star formation scheme based on
regions having the typical density of molecular cloud complexes. Supernovae-driven winds
become more effective in such mode, driving low angular momentum baryons outside the virial 
radius at high redshift and turning the dark matter cusp into a core.
Finally we show the first tidal stirring experiments adopting dwarfs formed in cosmological
simulations as initial conditions. The latter are gas dominated and have 
have turbulent thick gaseous and stellar disks disks that cannot develop strong bars,
yet they are efficiently heated into spheroids by tidal shocks.

\end{abstract}
\maketitle
%%-----------------------------
%%      your text
%%-----------------------------

\section{Introduction}

Dwarf galaxies are key to understand galaxy formation as well as the nature of dark matter.
Gas-rich dwarfs have slowly rising rotation curves that are reproduced better with dark matter
halos having a central core rather than the cuspy profiles predicted by CDM (see the review
by de Blok 2010 and Oh et al. 2008 for the recent results of the THINGS survey).
These galaxies also have nearly exponential profiles and no central stellar bulges. Cosmological
simulations of galaxy formation, instead, always produce galaxies with prominent bulges and
steep rotation curves, irrespective of the mass scale (Mayer et al. 2008).
Dwarf spheroidals (dSphs) in the Local Group are the faintest galaxies known. Including the
ultra-faint dwarf spheroidals discovered in the last few years, they span
luminosities in the range $-3 < M_B < -14$.
At variance with gas-rich dwarfs, they are gas poor
and have pressure supported stellar components (Mateo 1998). Among them some
stopped forming stars about 10 Gyr ago and other have extended star formation
histories (Hernandez et al. 2000; Coleman \& de Jong 2008; Orban et al. 2008).
They are typically clustered around the largest galaxy in a
group, although a few of them are found also at significantly larger distances
from the primary galaxy (Mateo 1998).
These properties of dSphs, while best studied and known in the Local Group
due to its proximity, are also typical of this class of galaxies in nearby groups and clusters.
Chiboucas, Karachentsev \& Tully (2009) have recently uncovered a population of dSphs in the
M81 group having a range of luminosities comparable to the LG dwarfs and which share
with the latter the same scaling relations between fundamental structural properties, such
as the relation between luminosity and effective radius. 
Recent studies of early-type dwarfs in 
clusters (Misgeld, Hilker
\& Mieske 2009) show that there
is overlap in the structural properties between the faint spheroidals in clusters and the classic 
dSphs in the LG,
as shown by the size-luminosity and the surface brightness-luminosity
relations. Strong similarities between dSphs in the LG and in clusters are also found by Penny et 
al. (2009) in their study of the Perseus cluster.
Hence, LG dSphs can be considered as a representative laboratory for the study of dSphs in the 
Universe.
In this paper we review the substantial progress made over the last years in two areas. First,
we summarize results of a scenario in which tidal stirring and ram pressure stripping with the 
halos of primary galaxies drive the transformation from gas-rich dwarfs to dwarf spheroidals over 
a few orbits of the dwarf satellites (5-10 Gyr). The we discuss the results of the first high 
resolution cosmological simulations of dwarf galaxy formation capable of obtaining realistic
rotation curves and suppress bulge formation. Finally we conclude with outlining work in progress 
that aims at combining both scenarios.

\begin{figure}
%\hskip=2truecm
%\epsfxsize=2truecm
\center
\includegraphics[height=5.6in,width=5.6in,angle=0]{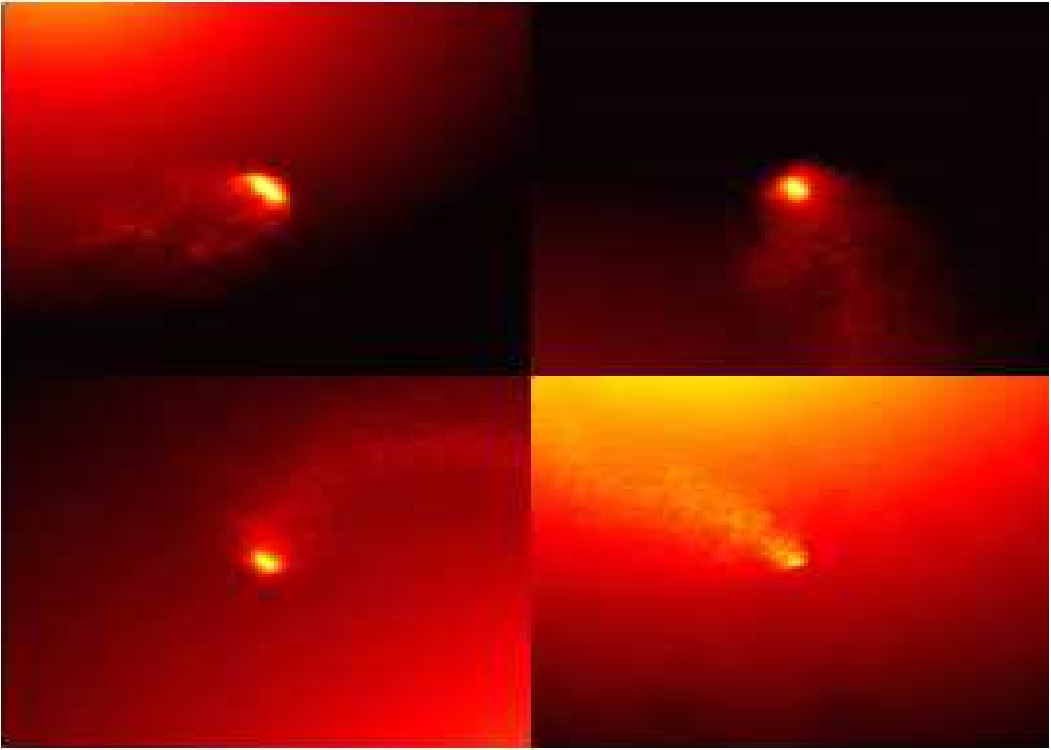}
\caption{Evolution of the gas-dominated disky dwarf model studied in
Mayer et al. (2007), which produces a dark matter dominated dwarf
resembling Draco after 10 Gyr of evolution. This N-Body + SPH simulation employed
millions of dark matter, gas and star particles. It included tidal mass loss due to a live
Milky Way halo, ram pressure stripping in a tenuous gaseous halo,
radiative cooling and a time-varying cosmic ionizing UV background consistent with
the Haardt \& Madau (2000) model. The snapshots show the color
coded logarithmic density maps (the brighter the color the higher the
density, with densities in the range $10^{-32} - 10^{-23}$ g cm$-3$)
for the first 2.5 Gyr of evolution. Boxes are 30 kpc on a side.
The dwarf begins falling
into the Milky Way halo on a typical eccentric cosmological orbit (apo/per$=5$ at the
beginning of the simulation, which corresponds to $z=2$).
At the first pericenter passage ($R_{peri} = 30$ kpc)  a prominent ram pressure tail is evident
(top left), and once the dwarf comes to first apocenter (top right) it
has lost already more than half of its gas. As it begins the second orbit
ram pressure stripping continues to remove gas (bottom left),
until all gas is stripped at second pericenter (bottom right).}
\end{figure}

\section{Tidal stirring, ram pressure stripping and the timing of infall into the 
primary: the transformation of dIrrs into dSphs}

Mayer et al. (2001a,b), Klimentowski et al. (2007;2009a) and Lokas et al. (2010) have shown that low surface 
brightess disky dwarfs are turned into
pressure supported spheroidal systems as a result of repeated tidal shocks at pericenter
passages as they orbit within the primary halo.
 The timescale
of the transformation is a few orbital times (several Gyr). The
mechanism behind the transformation is tidally induced
non-axisymmetric instabilities of stellar disks  combined with impulsive tidal heating
of the stellar distribution. The typical orbits considered are consistent with
those of satellites in cosmological simulations, having an apocenter to pericenter rauo of 5-6.

We recall that the time-dependent tidal field
varies on a timescale proportional to the orbital timescale of the galaxy.
In simulations of disky dwarfs on eccentric orbits interacting with a Milky Way-sized halo the 
following
sequence of events is typically observed.
First, tidal shocks induce strong bar
instabilities in otherwise stable, light disks resembling those of
present-day dIrrs. Second, the bar buckles due to the amplification
of vertical bending modes and turns into a spheroidal component in disks
with relatively high stellar surface density (Mayer et al. (2001a,b), or else subsequent shocks 
destroy the centrophilic
orbits supporting the bar which then loses its elongation and heats up into
a more  isotropic diffuse spheroid (Mayer et al. 2007; Klimentowski
et al. 2009a).
The second channel for the transformation is favoured in
systems with lower mass, lower surface density disks; for these tidal heating
is particularly efficient because the dynamical response of the stellar system is
impulsive rather than adiabatic.
Kazantzidis et al. (2011) have shown, by exploring a wide parameter space, that both
the pericenter distance and the orbital time are the key parameters that set
the efficiency of the transformation (see the paper for a detailed discussion).
The stars lose angular momentum as the bar instability arises, transferring it to the outer 
regions, that are mostly tidally stripped. The result is thus a system with a low $v/\sigma$,
below $0.5$, consistent with dSphs.
Mayer et al. (2006) studied the combined effect of ram pressure and tidal
stripping, showing how only when both are considered simultaneously is the original
gas content of the dwarf irregular progenitor removed in a few Gyr (a couple of orbits).
This result was obtained assuming that a primary galaxy such as the Milky Way has 
hot gaseous halo with density a few times $10^-5$ atoms/cm$^3$ at 50 kpc, with a power-law 
profile, consistent with various observational constraints.

In Mayer et al. 2007 we built on the results of Mayer et al. (2001a,b) and Mayer et al. (2006) 
and proposed a coherent scenario that explains at the same time the origin of the
common properties of dSphs (low gas content, exponential profiles, low luminosity
and surface brightness, low angular momentum content) and their differences (different star 
formation
histories and mass-to-light ratios). In this model the key parameter is the epoch of accretion
onto the Milky Way and a key assumption is that the progenitors of present-day  classic dSphs 
were
not simply gas-rich but extremely gas dominated, consistent with what is found in most
present-day dIrrs in the Local Group and elsewhere
in the nearby Universe (Geha et al. 2006). The large gas fractions found in field dwarfs can be
understood in terms of a decreasing star formation efficiency towards decreasing galaxy mass.
Recently, the THINGS survey (Leroy et al. 2008) has confirmed that disky dwarfs have a star
formation efficiency, defined as the fraction of gas (atomic+molecular) that is converted into 
stars,
well below that of normal spirals. The low gas surface densities typically
found in dwarfs likely imply a low conversion efficiency between atomic hydrogen and the star 
forming
molecular hydrogen phase (Schaye 2004), probably explaining the low star formation efficiency. 
Such conversion can be even less efficient
in presence of the ionizing ultraviolet flux arising during reionization, which can dissociate
molecular hydrogen (Schaye 2004). Therefore for field dwarfs that were accreted by the Milky Way
or M31 at $z > 1$ the assumption of mostly gaseous baryonic disk is even more well grounded.

\begin{figure}
\vskip 13cm
%\vspace{6.cm}
{\includegraphics{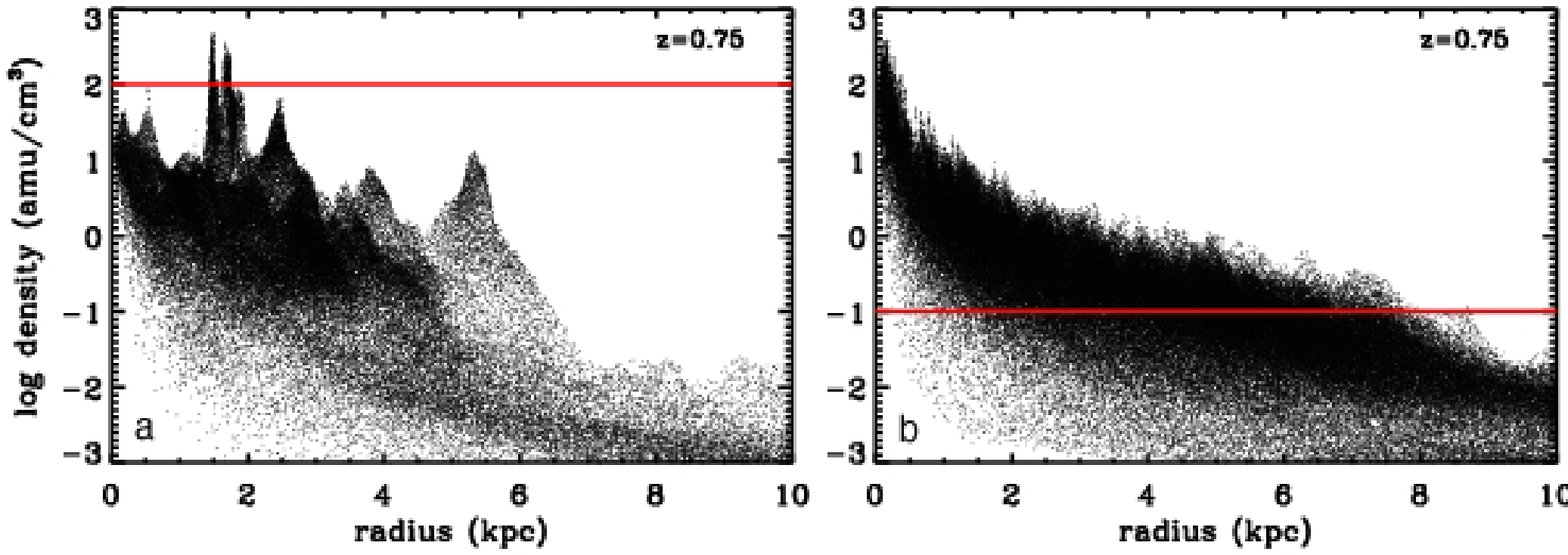}}
\caption[]{\small Properties of the gas distribution for different SF
  implementations. The local gas density measured around each SPH
  particle is plotted as a function of its radial distance from the
  galaxy center for runs DG1MR and DG1LT, which have identical force
  and mass resolution, but differ in the star formation density
  threshold.  Horizontal red lines mark the minimum gas density for SF
  in each run.  In both runs the SFR is $\propto$
  ${\rho_{gas}}^{1.5}$.  In the low threshold simulation, diffuse star
  formation in the inner regions continues unabated by feedback, as SN
  energy is more evenly distributed and is unable to originate major
  outflows.  Allowing star formation only in high density regions
  results in a complex, inhomogeneous ISM, even in the central regions
  and fast outflows that remove gas preferentially from the galaxy
  center.}
\label{dens_rad}
\end{figure}

The initial conditions, including the orbits of the satellites, were chosen
based on a hydrodynamical cosmological simulation of the formation of a Milky Way-type galaxy
(Governato et al. 2007).
We found that satellites that were accreted when the cosmic ionizing background was still
high, roughly before $z=1$, were completely ram pressure stripped of their gas in one to two 
pericenter
passsages (Figure 1).
As a result their star formation was truncated.
The effect of the ionizing radiation is to heat and ionize the gas, 
making it more
diffuse because of the increased pressure support, and to suppress star formation.
This, in turn, makes it easier to strip even from the
central regions of the dwarf, essentially having the same effect of a reduced binding energy.
Hence gas-rich dwarfs accreted when the UV radiation background was at its peak lost most of 
their baryonic
content because this was initially in gaseous form, thus naturally ending up with
a very low luminosity. While their baryonic content dropped orders
of magnitude below the cosmic mean as a result of gas stripping, their original central dark 
matter mass
in the central region around the surviving baryonic core
was largely preserved because dark matter is affected only by tides, not by ram pressure.
This automatically produced very high mass-to-light ratios, of order 100 (see Figure 2 in Mayer
et al. 2007).
We showed that all the final properties of such systems after 10 billion years of evolution, 
including
the stellar velocity dispersion profiles, resemble those of the classic strongly dark matter 
dominated
dwarfs such as Draco, Ursa Minor or And IX. Even the brightest among the ultra-faint dwarfs, that 
have velocity
dispersions $> 5$ km/s (e.g. Ursa Major) may be explained by this model. However, the very low
characteristic mass scale of most ultra-faint dwarfs suggests that other formation paths might 
indeed be
more likely.

Dwarfs that fell into halos of bright galaxies below $z=1$, when the cosmic
UV radiation dropped by more than an order of magnitude, retained some gas because tides
and ram pressure could not strip it completely, and underwent subsequent episodes of star 
formation
at pericenter passages due to bar-driven inflows and tidal compression (Mayer et al. 2001a,b).
These ended up in dSphs that are brighter for a given halo mass (or given central
stellar velocity dispersion) compared to the ones that were accreted earlier.
This is should be case with e.g. Fornax, Carina or Leo I. The two regimes of infall epochs
explain why Fornax and Draco have roughly the same halo peak circular velocities (and thus mass)
despite having a luminosity and mass-to-light ratio
that differs by about an order of magnitude. Likewise, Carina and Leo I, these being
prototypical cases of dSphs with an extended (episodic) star formation history, have
a luminosity comparable to Draco but a mass-to-light ratio 5-10 times lower than that of Draco
(Mateo 1998).

As a final remark, we note that Madau et al. (2008) argue that a very low efficiency of star 
formation,
corresponding to less than $0.1 \%$ of their total mass being converted into stars, would offer
a solution to the excess in number counts of subhalos with $V_c > 20$ km/s in dark matter-only 
simulations
(see also Koposov et al. 2009 for a similar interpretation).
Our model for the origin of Local Group dwarf spheroidals provides a clue to
why dSphs were so inefficient at producing a stellar component, thus pointing to
a solution of the substructure problem at the bright end of the mass function of
dwarf satellites of the Milky Way, which essentially contains all the classic dSphs.
Instead of relating the low efficiency of star
formation to photoevaporation and/or suppression of gas accretion by
the cosmic ultraviolet background,
we argue that it arose naturally from intrinsic low  star formation
efficiency in the progenitor gas-rich dwarfs (well below 1\% - see also Robertson \& Kravtsov 2008)
combined with copious gas stripping after they were
accreted in the potential of the primary galaxy. Our mechanism is absolutely general in 
hierarchical
structure formation and should thus apply to dwarf galaxy satellites of any galaxy group.
The combination of an intrinsic low star formation efficiency prior to infall with ram pressure
and tidal stripping can be thought of as an effective feedback mechanism alternative
to reionization and supernovae feedback.

\begin{figure}
\vskip 11.cm
{\includegraphics{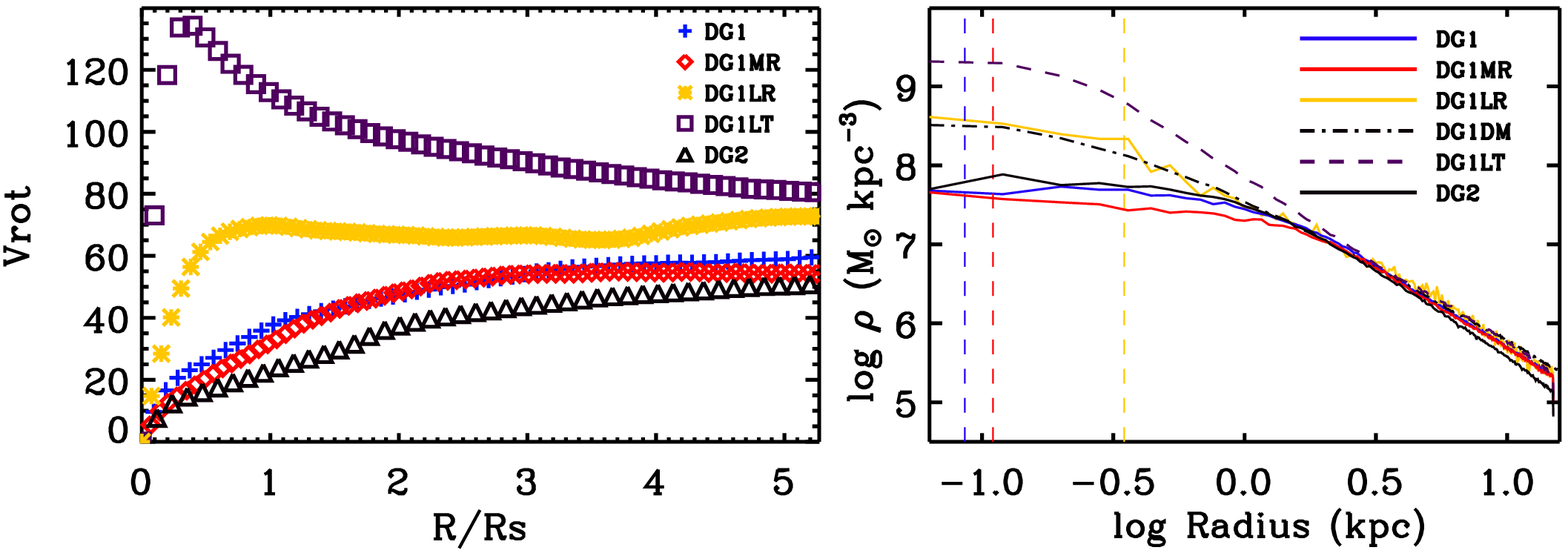}}
\caption[]{\small The rotation curve and DM radial distribution of the
  models described in \S3 and \S5, to show the effects of resolution
  and different SF recipes on the central mass distribution of
  simulated dwarf galaxies. The left panel shows the rotation curve,
  derived using the 3D potential of each galaxy and measured at z$=$0,
  for DG1 (blue crosses), DG1MR (red diamonds) and DG1LR, (yellow
  stars), plotted versus the disc scale length (1kpc for DG1, 0.5kpc
  for DG2). The three runs use the same gas density threshold for SF
  (100 atoms/cm$^3$), but MR and LR runs use only 40\% and 12.5\% of
  the particles of the reference run (with particle masses rescaled to
  the same total mass), and a softening respectively 1.33 and 4 times
  larger. Results have converged at the DG1MR resolution, while the LR
  run shows an excess of central material that is due to poor
  resolution causing artificial angular momentum loss (Mayer et al. 2008). The
   squares show DG1LT, where star formation is allowed in
  regions with a much lower local density (0.1 atoms/cm$^3$), again
  resulting in a much higher central mass density, due to the lack of
  outflows. This result demonstrates that the correct modeling of
  where SF is allowed to happen (namely only in gas with density
  comparable to that of real star forming regions) is crucial to
  obtain the results described in Governato et al. 2010 and summarized here. The rotation curve of
   galaxy DG2 (black triangles) shows the same shape as that of the DG1
  run.  Panel B uses a similar colour scheme and plots the DM density
  profile for the same runs. DG1 (blue solid), DG1MR (red dashed) show similar
  profiles, with a DM core of about 1kpc. Color coded vertical lines
  mark the force resolution for each run. The DM only run DG1DM (dot
  dashed), shows instead a cuspy profile down to the force resolution
  (red dashed vertical line as for DG1MR). The central density is
  about 10 times higher than in the runs with strong outflows. DG2 has
  a similar profile to DG1, while the lower resolution run or the run
  with diffuse SF have dense and cuspy DM profiles down to the force
  softening length.}
\label{rotcurve}
\end{figure}

\section{The formation of a realistic dwarf galaxy; inhomogeneous ISM and supernova-driven
outflows}

Here we focus on one important aspect of the recent paper Governato et al. (2010),
in which we showed for the first time the formation of a realistic dwarf
galaxy in a cosmological simulation of the concordance (WMAP5) cosmology.
We refer to the paper for the technical details of the simulations.
Several analytical and numerical papers have highlighted the necessity
of resolving a clumpy multi-phase ISM to achieve a realistic modeling
of energy deposition in the central regions of galaxies.  While using
a variety of arguments, these works suggest that only in a clumpy ISM
is it possible to a) transfer orbital energy from gas to the DM as
dense clumps sink through dynamical friction or resonant
coupling (Mashchenko et al. 2006,2008;Binney et al.2001;Mo \& Mao 2004, El-Zant et al. 2004) 
and have b) efficient gas
outflows (Maller \& Dekel 2002,. In turn, these outflows 1) suppress the formation
of stellar bulges by removing negative or low angular momentum
gas (Van den Bosch et al. 2001; Binney et al. 2001) and 2) make the central DM expand by suddenly
reducing the total enclosed mass and reducing the DM binding
energy (Navarro, Eke \& Frenk 1996; Mo \& Mao 2004; Read \& Gilmore 2005).  However, none of the above works has simultaneously
studied the formation of bulgeless galaxies and that of DM cores, even
if they both are crucial properties of small galaxies.

In order to achieve a multi phase ISM numerical works agree that a
minimal spatial resolution of about 100 pc is required, and that SF has
to be associated with dense regions with gas density ($\sim$ 100
amu/cm$^3$) (Robertson \& Kravtsov 2008; Saitoh et al. 2008).  Our simulations satisfy both conditions and
unify the many proposed models that focused on different aspects of
the problem to robustly demonstrate that energy transfer and
subsequent baryon removal are concurrent and effective to create
bulgeless galaxies with a shallow DM profile in a full cosmological
setting.

To illustrate the clumpiness of the ISM in our simulations, Figure~2
highlights the differences in the density distribution of the
interstellar medium between the simulations DG1MR and DG1LT (see Table 1), by
plotting the local gas density vs radius of each gas particle at a
representative epoch of z$=$0.75. These two runs have the same mass
and spatial resolution and adopt identical feedback schemes. They only
differ in the way regions where SF happens are selected.  In the
``high threshold'' runs (DG1,DG2 DG1MR and DG1LR) SF happens only in
regions above a high gas density threshold (100 amu/cm$^3$, the
horizontal red line in the left panel of Figure 2).  The density peaks
then correspond to isolated clumps of cold gas with masses and sizes
typical of SF regions.  The efficiency of SF, $\epsilon$SF, for these regions
must be increased from 0.05 (LT) to 0.1 (HT) in order to match the
observed normalization of SF density in local galaxies.  However, due
to the increased densities, at any given moment only a few regions are
actively forming stars.  These star forming regions get disrupted
after the first SNe go off and only a small fraction of gas has been
turned into stars. Feedback then  creates an ISM with cold filaments and
shells embedded in a warmer medium. This patchy distribution allows
the hot gas to leave the galaxy perpendicular to the disk plane at
velocities around 100 km/s.  Rather than developing a clumpy ISM as
in the ``high threshold'' case, SF in the ``low threshold'' scheme is
spatially diffuse (Figure~2).  This means that SN energy is more
evenly deposited onto the gas component, but less overall gas is
effected by SN feedback due to the low densities in the SF regions.
By monitoring where SN energy is deposited and where gas gets
substantially heated at high instantaneous rates, we verified that in
the ``high threshold'' case a larger mass of gas achieves temperatures
higher than the virial temperature (T$_{vir}$ $\sim$ 10$^5$ K) per
unit mass of stars formed than in the ``low threshold'' scheme.  Since
less mass is affected in the ``low threshold'' scenario, the outflows
are weak compared to the ``high threshold'' case.  By z$=$0 DG1LT has
formed ten times more stars, most of them in the central few kpcs,
causing strong adiabatic contraction of the DM component. Its light
profile is consistent with a $B/D$ ratio of $0.3$, typical of much more
massive galaxies and more concentrated than in real dwarfs.

We note that in the runs adopting the ``high threshold'' SF, feedback
produces winds that are comparable in strength to those happening in
real galaxies of similar mass. However, in our simulations the cold ISM is
still only moderately turbulent ($\sim$ 10 km/s at z=0), consistent
with observations (Walter et al. 2008), and the galaxies match the observed stellar
and baryonic Tully Fisher relation (Blanton et al. 2008), as the SF efficiency is
regulated to form an amount of stars similar to that of real dwarf
galaxies of similar rotation velocity, as demonstrated by the parallel
analysis done by the THINGS team to compare mock observations of DG1 and DG2
with dwarf galaxies in the THINGS survey (Oh et al. 2011).

The different ISM structure and, consequently, the different local strength of supernovae 
drive outflows, produces dramatic differences in the final mass distribution
and rotation curves (Figure 3). Only in the high resolution, high SF density threshold
runs the resulting rotation curve is slowly rising, corresponding to the absence
of a bulge and a dark matter profile that is cored rather than cuspy as a result
of impulsive heating of the dark matter from supernovae ouflows (Read \& Gilmore 2005) and transfer of
energy and angular momentum of gas clumps to the dark matter background (see
the Online Supplementary Information of Governato et al. (2010) and the detailed 
kinematical analysis performed in Oh et al. 2011 exactly as for the dwarfs in the THINGS
survey).

\section{Tidal stirring of thick, turbulent disky dwarfs forming in cosmological simulations}

In the previous sections we have summarized the tidal stirring scenario for the transformation of dwarf irregulars
into dwarf spheroidals and we have discussed the results of the first cosmological simulations describing t
he formation of realistic dwarf galaxies. The latter produce galaxies whose mass distribution resembles closely
 that of dwarf irregulars and late-type dwarfs of the THINGS sample (Oh et al. 2011), with faint low surface brightness 
disks and slowly rising rotation curves.  The shape of the rotation curve, which matches closely observed ones, is 
the result of three key factors; a realistic baryonic disk with a baryon fraction well below universal
and a low central density, the absence of  a bulge component and the fact that the dark halo is not cuspy,
 rather has a shallow inner power law density profile. Of these properties only the first two are satisfied by 
construction in the ICs normally used in tidal stirring works. At z=0 the galaxy has a peak circular velocity 
in the range 50-60 km/s, only slightly larger to that used in eg initial conditions of tidally stirred dwarfs that
can match the properties of classical dSphs such as Draco and Ursa Minor (Mayer et al 2007).
The dwarf galaxies produced in cosmological simulations, simlarly to real dIrrs, have thick stellar and gaseous
 disks (aspect ratio close to 3:1), a consequence of the turbulent motions triggered by supernovae winds 
and outflows in the gas and then inherited by the newly born stars. This is inevitable in scenario in which supernovae
outflows are so strong to eject more than half of the baryons and alter the slope  of the dark matter distribution
Sanchez-Janssen et al. (2010) show that dIrrs have normally thick disks. Low mass galaxies with circular velocities lower
than 50 km/s are expected to be thick as a natural result of the balance between thermal energy, gravitational
binding energy and rotational energy for standard halo spin parameters (Kaufmann et al. 2007).
Conversely, model galaxies for tidal stirring,  employ  a thin disk for both the stellar and  gaseous component
, with aspect ratio  10:1 (Mayer et al. 2001).
Gaseous disks can become much thicker as a result of the heating by the cosmic UV background
(Mayer et al. 2006), with aspect ratios even as low as 3:1 , but the inital stellar disk can only 
be thickened by tides, hence only after the interaction with the main galaxy has begun.

\begin{figure}
\vskip 13cm
%\vspace{6.cm}
{\includegraphics{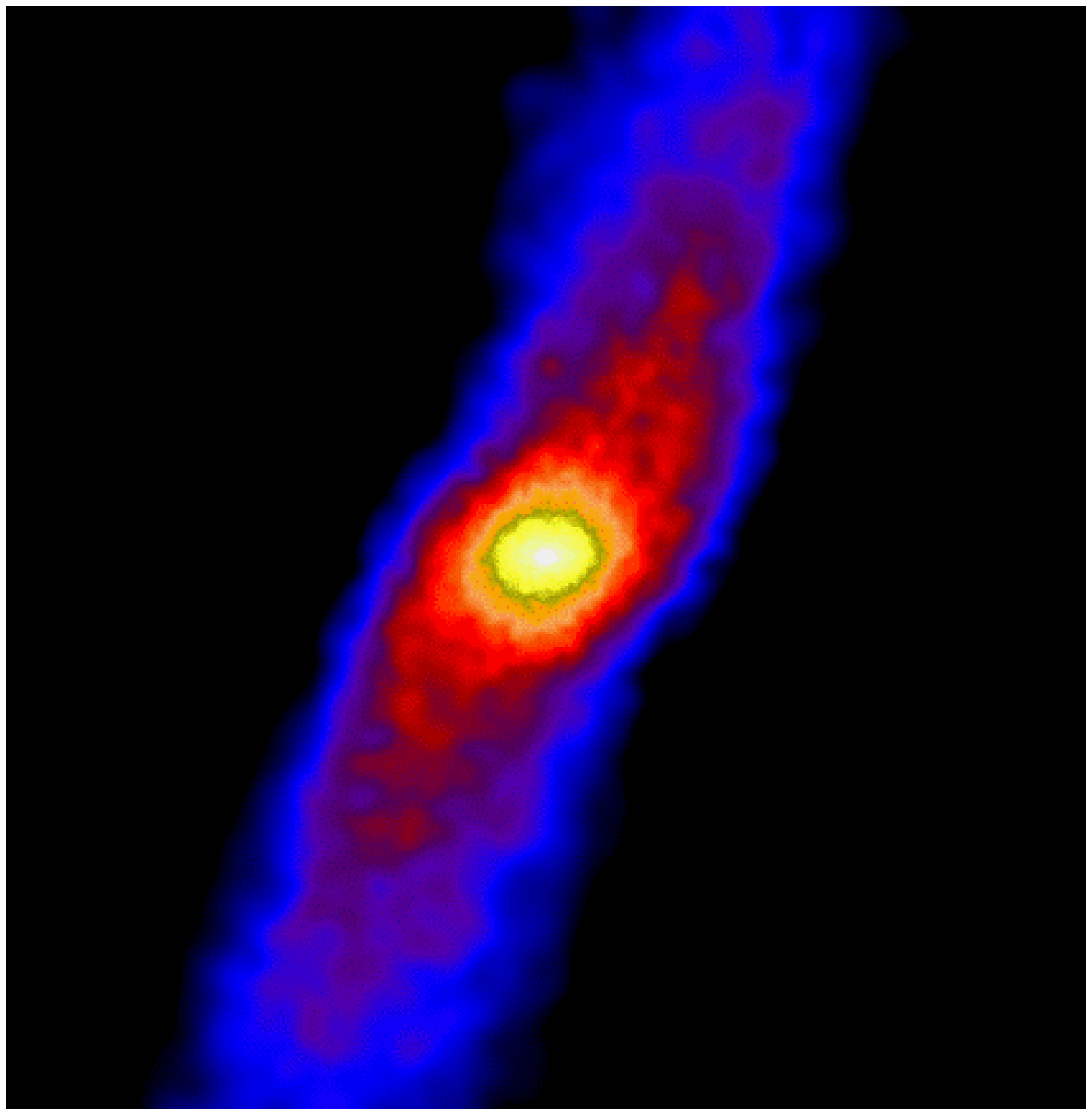}}
\caption[]{\small Evolved state of the cosmological dwarf DG1 extracted at z=1
 and placed on orbit inside a live Milky Way model. The color-coded density map
 of the stars seen face-on (perpendicular to the angular momentum vector) is shown. 
  The dwarf is displayed after first pericenter passage.
  The tidal distorion is evident but there is no sign of a strong bar}
\end{figure}

At comparable mass of the disk and halo, which determine the depth of the potential well and hence
the strength of the gravitational restoring force, a thicker disk will have a higher vertical stellar velocity 
dispersion. As a result, it will have a lower self-gravity. Perturbations of the stellar surface density field
such as the bar-like modes, should grow less efficiently. For isolated systems and weak perturbations modifications
of the standard Toomre analysis show that the critical Toomre Q parameter for axisymmetric modes is lower 
than unity (close to 0.67, see eg Nelson et al. 1998).
Here we show preliminary results of new tidal stirring simulations that employ directly 
the cosmological
dwarfs in the Governato et al. (2010) paper. These are the first hydrodynamical simulations that explore tidal 
interactions using an hybrid approach, namely
combining the very high resolution  proper of idealized binary interaction experiments with cosmological 
initial conditions for one of the two galaxies (the dwarf satellite in this case).
The DG2 dwarf is extracted at two different cosmic epochs during the simulations (z=2 and z=1) and implanted 
in the same multi-component model of the Milky Way used in the experiments of Mayer et al. (2006, 2007) and 
reviewed in section 2, which includes a hot  gaseous corona, and therefore allows to model the effect of ram pressure.   
The adopted eccentric orbit is similar to that chosen by Mayer et al. (2007).

The two different extraction epochs correspond effectively to two different dwarf models; not only the dwarf 
grows in mass from z=2 to z=1, but also changes in structure, 
becoming more rotationally supported owing to accretion of higher angular momentum material, acquiring a shallower 
halo density profile, yet maintaining a fairly high thickness.
Figure 5 shows the first results of the simulation adopting DG2 extracted at z=1, with the dwarf caught
soon after the first pericenter passage. At this time the dwarf
has a peak circular velocity of about 45 km/s, similar to that of the Mayer et al. (2007) ICs, 
it is gas-dominated and has already a slope shallower than NFW. Ram pressure stripping has already removed
the outer gas disk.
As shown in the Figure the disk is tidally distorted but there 
is no sign of the strong bar-like mode
ubiquitous in previous tidal stirring runs at the same stage. Yet the disk has been heated into a spheroid
and $v/\sigma$ is lower than 0.5 near the end of the second orbit, which is how far the run has progressed so far, 
having decreased by almost a factor of 4 relative to the initial state. It appears  that direct tidal heating is the culprit
behind the transformation rather than tidally induced non-axisymmetric instabilities. Kazantzidis et al. (2011), that 
have completed the largest parameter survey for tidally stirred dwarfs starting with equilibrium models,
find that disk thickness had little effect on the transformation. In the latter work the amplitude of the bar mode
decreases faster after the first orbit in the thicker disk case, but the bar is forming nevertheless, 
and the final state was extremely close to that of models with disks even three times thinner. The thick
disk case had a disk scale height about a third of the scale length, not far from the cosmological dwarf.
Therefore the reason why the bar does not form in the cosmological dwarf cannot be simply traced to the
thicker disk, rather other structural properties might play a role in combination to it. One important
difference is that the cosmological dwarf has a disk which is gas dominated rather than purely stellar
as in Kazantzidis et al. (2011), which further decreases the stellar disk surface density and its
response to the tidal perturbation. Mayer et al. (2007) considered gas dominated dwarfs but with initially thin stellar disks
(aspect ratio $\sim 0.1$); the bar was forming at first pericenter but was dissolving soon afterwards, and direct
tidal heating was then taking over. It is thus possible that a disk that is both gas dominated and thick the
response is so weak that the bar does not arise at all. A detailed analysis of the evolution of the Toomre
and swing amplification parameters will shed light on this. It will be performed in a a forthcoming paper.
Therefore, we expect this new work will confirm the tidal sirring scenario for dwarfs forming in a 
cosmological context, but will also show that non-axisymmetric instabilities are not as crucial as previously
found in driving the transformation.

\newpage

%\begin{table*}
\centering
\begin{tabular}{lcccccc}
\hline

Run & M$_i$  & $g-r$ & SFR  & R$_s$ $i$  & V$_{rot}$ & M$_{HI}$/L$_B$  \\
& &   &M$_{\odot}$/yr& kpc & km/s & $ $ \\
\hline
\\
DG1     & -16.8  & 0.52 & 0.01 & 0.9  & 56  & 1.2   \\
DG1MR   & -16.9  & 0.54 & 0.02 & 0.9  & 55  & 1.0   \\
DG1LR   & -18.7  & 0.33 & 0.22 & 0.9  & 62  & 0.64    \\
DG1LT   & -19.4  &  0.40 &  0.38 & 1.3  & 78  & 0.11   \\
DG2     & -15.9 &   0.46 & 0.02 & 0.5  & 54  & 2.8   \\
\hline
\end{tabular}

\bigskip

{Summary of the observable properties of the different dwarf runs. the SFR is in M$\odot$/yr, R$_s$ is the disc scale length,  V$_{rot}$ is measured using $t$
\label{parameters}
%\end{table*}

%\subsection{...}
%\subsection{...}
%...
%%-----------------------------
%%      your bibliography
%%-----------------------------
%\begin{thebibliography}{99}

\end{document}